\documentclass[twocolumn,trackchanges]{aastex62}
\usepackage[utf8]{inputenc}

\usepackage{xspace}

\newcommand{\dmu}{pc\,cm$^{-3}$}
\newcommand{\expnt}[2]{\ensuremath{#1 \times 10^{#2}}}   
\newcommand{\psra}{J0038\ensuremath{-}2501\xspace}
\newcommand{\psrb}{J1916\ensuremath{-}2939\xspace}
\newcommand{\psrc}{J1949+3426\xspace}
\newcommand{\psrd}{J2355+2246\xspace}
\newcommand{\swift}{\textit{Swift}}
\begin{document}

\title{The Green Bank North Celestial Cap Pulsar Survey. IV: Four New Timing Solutions}

\author[0000-0003-2822-616X]{R.~J.\ Aloisi}
\affiliation{Center for Gravitation, Cosmology, and Astrophysics,
  Department of Physics, University of Wisconsin-Milwaukee, PO Box
  413, Milwaukee, WI, 53201, USA}
\author{A.\ Cruz}
\affiliation{Center for Gravitation, Cosmology, and Astrophysics,
  Department of Physics, University of Wisconsin-Milwaukee, PO Box
  413, Milwaukee, WI, 53201, USA}
\author{L.\ Daniels}
\affiliation{Center for Gravitation, Cosmology, and Astrophysics,
  Department of Physics, University of Wisconsin-Milwaukee, PO Box
  413, Milwaukee, WI, 53201, USA}
\author{N.\ Meyers}
\affiliation{Center for Gravitation, Cosmology, and Astrophysics,
  Department of Physics, University of Wisconsin-Milwaukee, PO Box
  413, Milwaukee, WI, 53201, USA}
\author{R.\ Roekle}
\affiliation{Center for Gravitation, Cosmology, and Astrophysics,
  Department of Physics, University of Wisconsin-Milwaukee, PO Box
  413, Milwaukee, WI, 53201, USA}
\author{A.\ Schuett}
\affiliation{Center for Gravitation, Cosmology, and Astrophysics,
  Department of Physics, University of Wisconsin-Milwaukee, PO Box
  413, Milwaukee, WI, 53201, USA}
\author[0000-0002-1075-3837]{J.~K.\ Swiggum}
\affiliation{Center for Gravitation, Cosmology, and Astrophysics,
  Department of Physics, University of Wisconsin-Milwaukee, PO Box
  413, Milwaukee, WI, 53201, USA}
\author[0000-0002-2185-1790]{M.~E.\ DeCesar}
\affiliation{Department of Physics, 730 High St., Lafayette College, Easton, PA, 18042, USA}
\author[0000-0001-6295-2881]{D.~L.\ Kaplan}
\affiliation{Center for Gravitation, Cosmology, and Astrophysics,
  Department of Physics, University of Wisconsin-Milwaukee, PO Box
  413, Milwaukee, WI, 53201, USA}
\author[0000-0001-5229-7430]{R.~S.\ Lynch}
\affiliation{Green Bank Observatory, PO Box 2, Green Bank, WV 24494, USA}
\affiliation{Center for Gravitational Waves and Cosmology, Department of Physics and Astronomy, West Virginia University, White Hall, Box 6315, Morgantown, WV 26506, USA}
\author[0000-0002-7261-594X]{K.\ Stovall}
\affiliation{National Radio Astronomy Observatory, 1003 Lopezville Rd., Socorro, NM, 87801, USA}
\author[0000-0002-2034-2986]{Lina Levin}
\affiliation{Jodrell Bank Centre for Astrophysics, School of Physics
  and Astronomy, The University of Manchester, Manchester, M13 9PL,
  UK}
\author[0000-0003-0638-3340]{A.~M.\ Archibald}
\affiliation{U. Amsterdam, the Netherlands Institute for Radio Astronomy,
  Postbus 2, 7990 AA, Dwingeloo, The Netherlands}
\author{S.\ Banaszak}
\affiliation{Center for Gravitation, Cosmology, and Astrophysics,
  Department of Physics, University of Wisconsin-Milwaukee, PO Box
  413, Milwaukee, WI, 53201, USA}
\author{C.~M.\ Biwer}
\affiliation{Department of Physics, Syracuse University, Syracuse, NY,
  13244, USA}
\author{J.\ Boyles}
\affiliation{Department of Physics and Astronomy, Western Kentucky
  University, 1906 College Heights Blvd., Bowling Green, KY,
  42101, USA}
\author[0000-0002-3426-7606]{P.\ Chawla}
\affiliation{Department of Physics \& McGill Space Institute, McGill
  University, 3600 University Street, Montreal, QC, H3A 2T8, Canada}
\author{L.~P.\ Dartez}
\affiliation{Center for Advanced Radio Astronomy, University of Texas
  Rio Grande Valley, 1 W.\ University Blvd., Brownsville, TX,
  78520, USA}
\author[0000-0003-3222-1302]{B.\ Cui}
\affiliation{Center for Gravitational Waves and Cosmology, Department
  of Physics and Astronomy, West Virginia University, White Hall, Box
  6315, Morgantown, WV 26506, USA}
\author{D.~F.\ Day}
\affiliation{Center for Gravitation, Cosmology, and Astrophysics,
  Department of Physics, University of Wisconsin-Milwaukee, PO Box
  413, Milwaukee, WI, 53201, USA}
\author{A.~J.\ Ford}
\affiliation{Center for Advanced Radio Astronomy, University of Texas
  Rio Grande Valley, 1 W.\ University Blvd., Brownsville, TX,
  78520, USA}
\author{J.\ Flanigan}
\affiliation{Center for Gravitation, Cosmology, and Astrophysics,
  Department of Physics, University of Wisconsin-Milwaukee, PO Box
  413, Milwaukee, WI, 53201, USA}
\author{E.\ Fonseca}
\affiliation{Department of Physics \& McGill Space Institute, McGill
  University, 3600 University Street, Montreal, QC, H3A 2T8, Canada}
\author[0000-0003-2317-1446]{J.~W.~T.\ Hessels}
\affiliation{ASTRON, the Netherlands Institute for Radio Astronomy,
  Postbus 2, 7990 AA, Dwingeloo, The Netherlands}
\affiliation{Anton Pannekoek Institute for Astronomy, University of
  Amsterdam, Science Park 904, 1098 XH, Amsterdam, The Netherlands}
\author{J.\ Hinojosa}
\affiliation{Center for Advanced Radio Astronomy, University of Texas
  Rio Grande Valley, 1 W.\ University Blvd., Brownsville, TX,
  78520, USA}
\author{C.\ Karako-Argaman}
\affiliation{Department of Physics \& McGill Space Institute, McGill
  University, 3600 University Street, Montreal, QC, H3A 2T8, Canada}
\author[0000-0001-9345-0307]{V.~M.\ Kaspi}
\affiliation{Department of Physics \& McGill Space Institute, McGill
  University, 3600 University Street, Montreal, QC, H3A 2T8, Canada}
\author[0000-0001-8864-7471]{V.~I.\ Kondratiev}
\affiliation{ASTRON, the Netherlands Institute for Radio Astronomy,
  Postbus 2, 7990 AA, Dwingeloo, The Netherlands}
\affiliation{Astro Space Center of the Lebedev Physical Institute,
  Profsoyuznaya str. 84/32, Moscow, 117997, Russia}
\author{S.\ Leake}
\affiliation{Center for Advanced Radio Astronomy, University of Texas
  Rio Grande Valley, 1 W.\ University Blvd., Brownsville, TX,
  78520, USA}
\author{G.\ Lunsford}
\affiliation{Center for Advanced Radio Astronomy, University of Texas
  Rio Grande Valley, 1 W.\ University Blvd., Brownsville, TX,
  78520, USA}
\author{J.~G.\ Martinez}
\affiliation{Max-Planck-Institut f\"{u} Radioastronomie, Auf dem
  H\"{u}gel 69, D-53121, Bonn, Germany}
\author{A.\ Mata}
\affiliation{Center for Advanced Radio Astronomy, University of Texas
  Rio Grande Valley, 1 W.\ University Blvd., Brownsville, TX,
  78520, USA}
\author[0000-0001-7697-7422]{M.~A.\ McLaughlin}
\affiliation{Department of Physics and Astronomy, West Virginia University, Morgantown, WV, 26501, USA and the Center for Gravitational Waves and Cosmology, Chestnut Ridge Research Building, Morgantown, WV, 26505, USA}
\author{H.\ Al Noori}
\affiliation{New York University Abu Dhabi, Abu Dhabi, UAE}
\author[0000-0001-5799-9714]{S.~M.\ Ransom}
\affiliation{National Radio Astronomy Observatory, 520 Edgemont Road,
Charlottesville, VA, 23903, USA}
\author[0000-0002-9396-9720]{M~.S.~E.\ Roberts}
\affiliation{New York University Abu Dhabi, Abu Dhabi, UAE}
\affiliation{Eureka Scientific, Inc., 2452 Delmer St., Suite 100,
  Oakland, CA, 94602, USA}
\author{M.~D.\ Rohr}
\affiliation{Center for Gravitation, Cosmology, and Astrophysics,
  Department of Physics, University of Wisconsin-Milwaukee, PO Box
  413, Milwaukee, WI, 53201, USA}
\author[0000-0002-1607-6646]{X.\ Siemens}
\affiliation{Center for Gravitation, Cosmology, and Astrophysics,
  Department of Physics, University of Wisconsin-Milwaukee, PO Box
  413, Milwaukee, WI, 53201, USA}
\author[0000-0002-6730-3298]{R.\ Spiewak}
\affiliation{Centre for Astrophysics and Supercomputing, Siwnburne
  University of Technology, PO Box 218, Hawthorn, VIC 3122, Australia}
\affiliation{Center for Gravitation, Cosmology, and Astrophysics,
  Department of Physics, University of Wisconsin-Milwaukee, PO Box
  413, Milwaukee, WI, 53201, USA}
\author[0000-0001-9784-8670]{I.~H.\ Stairs}
\affiliation{Department of Physics and Astronomy, University of
  British Columbia, 6224 Agriculture Rd., Vancouver, BC, V6T 1Z1,
  Canada}
\author[0000-0001-8503-6958]{J.\ van Leeuwen}
\affiliation{ASTRON, the Netherlands Institute for Radio Astronomy,
  Postbus 2, 7990 AA, Dwingeloo, The Netherlands}
\affiliation{Anton Pannekoek Institute for Astronomy, University of
  Amsterdam, Science Park 904, 1098 XH, Amsterdam, The Netherlands}
\author{A.~N.\ Walker}
\affiliation{Center for Gravitation, Cosmology, and Astrophysics,
  Department of Physics, University of Wisconsin-Milwaukee, PO Box
  413, Milwaukee, WI, 53201, USA}
\author{B.~L.\ Wells}
\affiliation{Department of Atmospheric Sciences, Colorado State
  University, 3915 W.\ Laporte Ave., Fort Collins, CO, 80523, USA}
\affiliation{Center for Gravitation, Cosmology, and Astrophysics,
  Department of Physics, University of Wisconsin-Milwaukee, PO Box
  413, Milwaukee, WI, 53201, USA}

\correspondingauthor{J.~K.~Swiggum}
\email{swiggumj@uwm.edu}

\begin{abstract}
We present timing solutions for four pulsars discovered in the Green Bank Northern Celestial Cap (GBNCC) survey.  
All four pulsars are isolated with spin periods between 0.26\,s and 1.84\,s.  PSR~\psra\ has a 0.26\,s period and a period derivative of $\expnt{7.6}{-19}\,{\rm s\,s}^{-1}$, which is unusually low for isolated pulsars with similar periods.  This low period derivative may be simply an extreme value for an isolated pulsar or it could indicate an unusual evolution path for PSR~\psra, such as a disrupted recycled pulsar (DRP) from a binary system or an orphaned central compact object (CCO).  Correcting the observed spin-down rate for the Shklovskii effect suggests that  this pulsar may have an unusually low space velocity, which is consistent with expectations for DRPs.  There is no X-ray emission detected from PSR~\psra\ in an archival \swift\ observation, which suggests that it is not a young orphaned CCO.  The high dispersion measure of PSR J1949+3426 suggests a distance of 12.3 kpc. This distance indicates that PSR J1949+3426 is among the most distant 7\% of Galactic field pulsars, and is one of the most luminous pulsars.
\end{abstract}


\section{Introduction}
The Green Bank North Celestial Cap (GBNCC) survey (\citealt{stovall14,lynch18,Kawash2018}) is searching for pulsars and transient radio signals at 350\,MHz in the declination ($\delta$) range available to the Robert C.\ Byrd Green Bank Telescope (GBT), $\delta > -40\degr$.  Scientific objectives for the GBNCC survey include characterization of the Galactic pulsar population as well as finding high precision millisecond pulsars (MSPs) suitable for inclusion in a pulsar timing array (PTA), which will enable the detection of nanohertz-frequency gravitational waves \citep[GWs; e.g.][]{abb+18}. 
By surveying the entire sky at low frequencies we are especially sensitive to nearby, low-luminosity and/or steep-spectrum pulsars; see \citet{stovall14} for further comparison of GBNCC's sensitivity with other pulsar surveys.  Note that the sensitivity of GBNCC also allows detection of more distant, higher luminosity pulsars as evidenced in the detection of \psrc reported in this paper.
As of 2018, GBNCC survey observations are approximately 80 percent complete, with the data collection expected to conclude by 2020.  As of 2018 October, 161 pulsars, including 20 millisecond pulsars have been discovered by the GBNCC survey.  
Initial GBNCC survey results were reported in \citet{stovall14}, while \citet{Kawash2018} discussed timing results for 10 pulsars and \citet{lynch18} reported timing results for an additional 45 pulsars.
This paper reports results from analysis of timing observations of four pulsars discovered in the GBNCC survey.
The University of Wisconsin--Milwaukee provided an opportunity for undergraduate students to participate in course-based research by processing data from observations on the four pulsars to develop timing solutions and to characterize the pulsars based on their properties\footnote{The pulsar timing analysis presented here is the culmination of efforts by the six lead authors, who participated in a First Year Research Experiences (FYRE) course held at the University of Wisconsin--Milwaukee during the Fall semester of 2017, {\it PHYS 194: {Clocking} Dead Stars with Radio Telescopes.}}.
In the discussion below, we give quantities and distances computed using both the Galactic electron density model of \citet[][YMW16]{Yao2017} and that of \citet[][NE2001]{ne2001}.

\section{Observations \& Timing Analysis}\label{sec:obs}
The discovery observations for the new discoveries presented here took place between 2011 and 2015 and used the GBT operating at a center frequency of 350\,MHz and nominal bandwidth of 100\,MHz, with dwell times of 120\,s; see \citet{stovall14} for a description of the methodology.  

The search processing took place on a computer cluster operated by Compute Canada, with candidates analyzed via the CyberSKA interface\footnote{\url{https://ca.cyberska.org/}}.
The timing observations for the four pulsars presented in this paper used the same center frequency and bandwidth, with typical durations of 3.5--6\,min.  The Green Bank Ultimate Pulsar Processing Instrument (GUPPI; \citealt{duplain08}) was used for both discovery and timing observations to record data every 81.92\,$\mu$s with 4096 frequency channels. 
Data were processed using {\tt PRESTO} \citep{rem+02} for initial spin period refinement, then {\tt PSRCHIVE} \citep{hotan04,vanStraten2012} to process individual timing scans and calculate times of arrival (TOAs).

\begin{figure}
    \includegraphics[width=1.0\columnwidth]{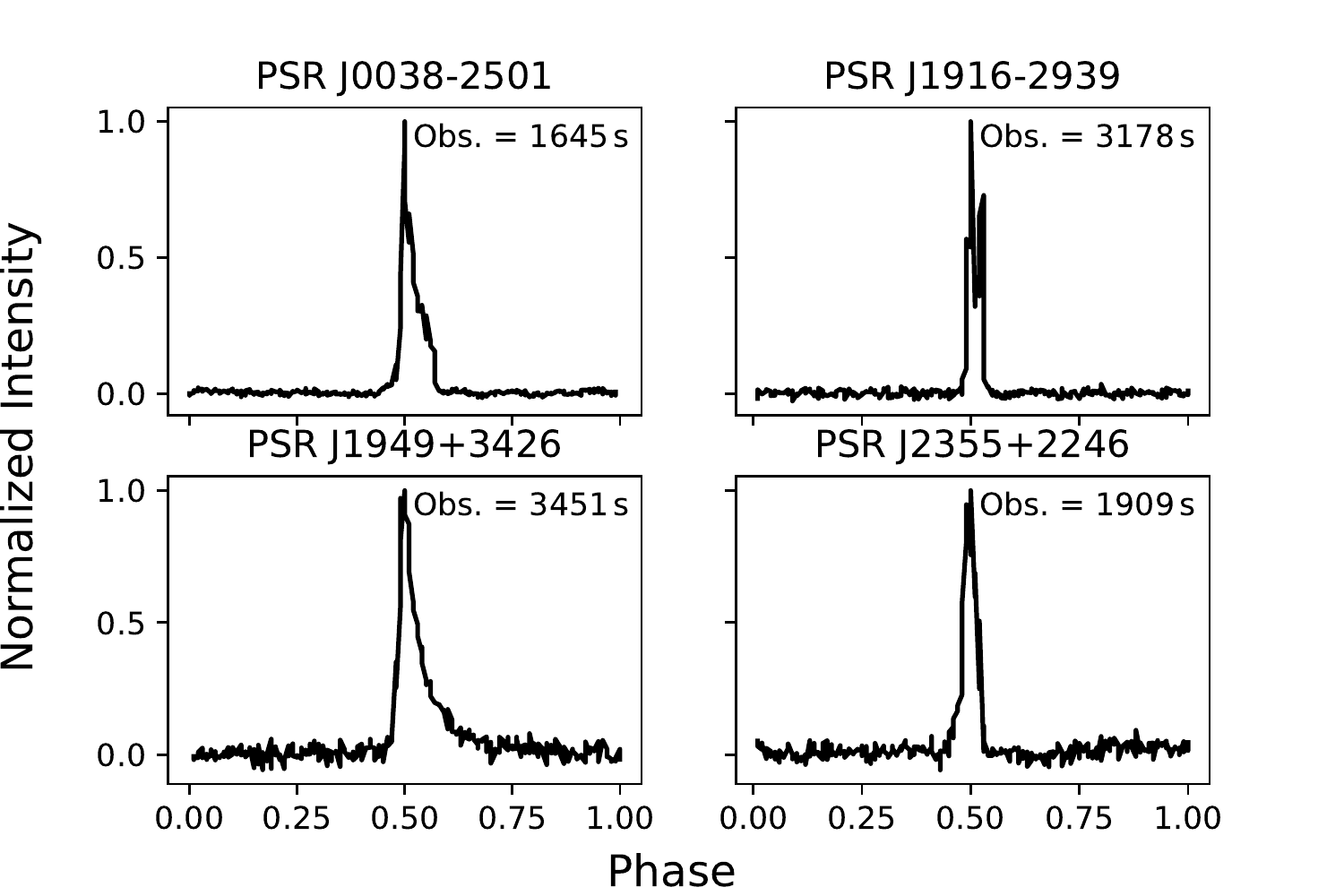}
	\caption{Normalized pulse profiles for observations at 350\,MHz of the four pulsars timed here.  We show PSRs~\psra\ (upper left), \psrb\ (upper right), \psrc\ (lower left), and \psrd\ (lower right). The total amount of observation time is noted in the upper right-hand corner for each summed profile.}
    \label{fig:profiles}
\end{figure}

\begin{figure}
    \includegraphics[width=1.0\columnwidth]{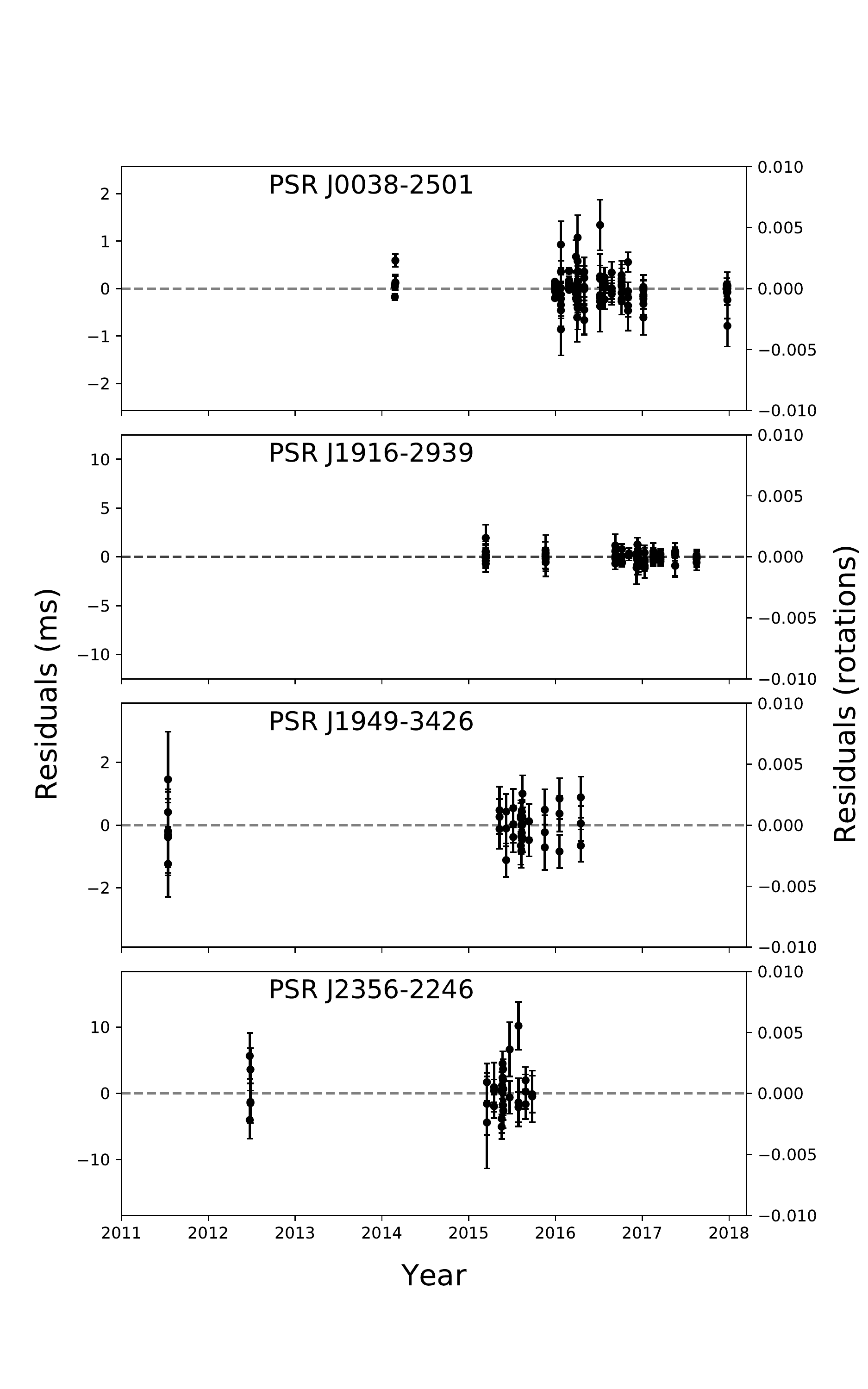}
	\caption{TOA residuals for three sub-integrations per epoch and one to three frequency sub-bands for each of the four pulsars timed here.   We show (top to bottom) PSRs~\psra, \psrb, \psrc, and \psrd.  The dense timing observations are apparent, with relatively large gaps that connect back to the discovery observations.  Error bars reflect 1-$\sigma$ uncertainties on TOAs.}
    \label{fig:Residuals}
\end{figure}

An ephemeris was created to save the preliminary timing parameters.  Using the dispersion measure (DM) found from the discovery observations, the  files from all timing observations were averaged from 4096 to 256 frequency channels using {\tt pam} and each file was examined using {\tt pazi} to remove radio frequency interference (RFI).  Figure \ref{fig:profiles} shows the composite profiles based on all timing observations for each pulsar.  Standard profiles were created for each pulsar using {\tt paas} on files with a high signal-to-noise ratio.  All timing files were then averaged in frequency again using {\tt pam} to one or more frequencies and three sub-integrations prior to using {\tt pat} and the standard profiles to generate TOAs; the number of frequency sub-bands was one for observations with relatively low signal-to-noise ratio and two or three for observations with higher signal-to-noise ratio.
Fitting the TOAs was performed with \texttt{TEMPO2} \citep{Manchester2015, hobbs06}, finding a  timing solution including spin period ($P$), period derivative ($\dot{P}$), and position.  We also included DM as a free parameter for pulsars for which TOAs were available at multiple frequencies.  Where not available from {\tt TEMPO2}, DM errors were determined using the {\tt PSRCHIVE} command {\tt pdmp}. Parameter uncertainties quoted in Table \ref{tab:spin} are 1-$\sigma$ uncertainties on measured {\tt TEMPO2} fit parameters, but a global multiplicative error factor (EFAC) has been applied to each TOA error such that the resulting reduced $\chi^2$ value is one after fitting.  Discovery observations were included in the timing analysis for each pulsar after similar processing using  {\tt PRESTO}.  After fitting model parameters, the profiles for TOAs with relatively large residuals were each examined visually to determine whether each was a significant detection. TOAs with no clear detection were deleted prior to the final model fit.  The final TOA residuals are plotted in Figure~\ref{fig:Residuals}.  We confirmed that the discovery TOAs could be reliably phase-connected with the timing TOAs by ensuring that the phase uncertainty extrapolated to the time of the discovery observations was much less than 0.1\,cycle.

Taken together, the data span of the combined discovery and timing data-sets are at least two years for each pulsar, so that covariances between spin-down and position are minimized.  The final timing models are given in Table~\ref{tab:spin}.

\begin{deluxetable*}{lcccc}
\tablewidth{0pt}
\tablecaption{Timing Solutions and Derived Parameters}
\tablehead{\colhead{Parameter} & \colhead{\object[PSR J0038-2501]{J0038$-$2501}} & \colhead{\object[PSR J1916-2939]{J1916$-$2939}} & \colhead{\object[PSR J1949+3426]{J1949+3426}} & \colhead{\object[PSR J2355+2246]{J2355+2246}}}
\startdata
Right Ascension (J2000) & $00^{\rm h}38^{\rm m}10\fs264(10)$ & $19^{\rm h}16^{\rm m}32\fs701(5)$ & $19^{\rm h}49^{\rm m}13\fs671(6)$ & $23^{\rm h}55^{\rm m}49\fs8(3)$ \\
Declination (J2000) & $-25\degr01\arcmin30\farcs73(2)$ & $-29\degr39\arcmin27\farcs8(3)$ & $+34\degr26\arcmin33\farcs89(8)$ & $+22\degr46\arcmin17(8)\arcsec$ \\
Galactic Longitude (deg) & 67.42 & 8.25 & 69.72 & 106.53 \\
Galactic Latitude (deg) & $-86.35$ & $-18.07$ & $4.29$ & $-38.32$ \\
Dispersion Measure (\dmu) & 5.710(3) & 38.34(11) & 228.0(3) & 23.1(7) \\
NE2001 Distance (kpc) & 0.32 & 1.2 & 9.8 & 1.2 \\
YMW16 Distance (kpc) & 0.60 & 1.6 & 12.3 & 2.2 \\
Spin Period (s) & 0.2569264575329(17) & 1.248616964290(3) & 0.3885391675859(14) & 1.8409859072(3) \\
Period Derivative ($10^{-17}\,{\rm s\,s}^{-1}$) & 0.0760(6) & 124.41(4) & 20.219(8) & 378(4) \\
Epoch (MJD) & 57474 & 57346 & 56921 & 57102 \\
Span of Timing Data (MJD) & 56774 -- 58175 & 56901 -- 57791 & 56051 -- 57791 & 56477 -- 57666 \\
Number of TOAs & 88 & 91 & 42 & 34 \\
RMS Fit Residual ($\mu$s) & 120 & 422 & 494 & 3061 \\
EFAC & 1.3 & 0.81 & 0.91 & 1.5 \\
Characteristic Age (Myr) & 5400 & 16 & 30 & 7.7 \\
Surface Magnetic Field ($10^9$\,G) & 14 & 1300 & 280 & 2700 \\
Spin-down Luminosity ($10^{30}\,{\rm ergs\,s}^{-1}$) & 1.8 & 25 & 140 & 9.6 \\
Signal to Noise & 81 & 18 & 23 & 12 \\
Pulse Width, W10 (s) & 0.015 & 0.061 & 0.030 & 0.115 \\
$T_{\rm sky}$ (K) & 27 & 70 & 77 & 28 \\
$T_{\rm sys}$ (K) & 73 & 116 & 123 & 74 \\
$S_{350}$ (mJy) & 3.7 & 2.2 &  3.7 & 0.9 \\
$L_{350}$ (mJy\,kpc$^2$) & 1.3 & 6.4 & 570 & 4.1 \\
\enddata
    \tablecomments{Numbers in parentheses are the 1-$\sigma$ errors in the last digit quoted after scaling TOA uncertainties by EFAC.  The signal to noise values are for discovery observations after RFI was removed.  Distance is calculated from DM using both the NE2001 \citep{ne2001} and YMW16 \citep{Yao2017} Galactic electron density models. Sky temperatures are calculated using \citet{Oliveira2008}. The flux densities and pseudo-luminosities are calculated using Equation \ref{eq:radiometer}.  The solar system ephemeris used was DE430.  The time scale used was TT(TAI).}
\label{tab:spin}
\end{deluxetable*}

\begin{figure}
\includegraphics[width=1.\columnwidth]{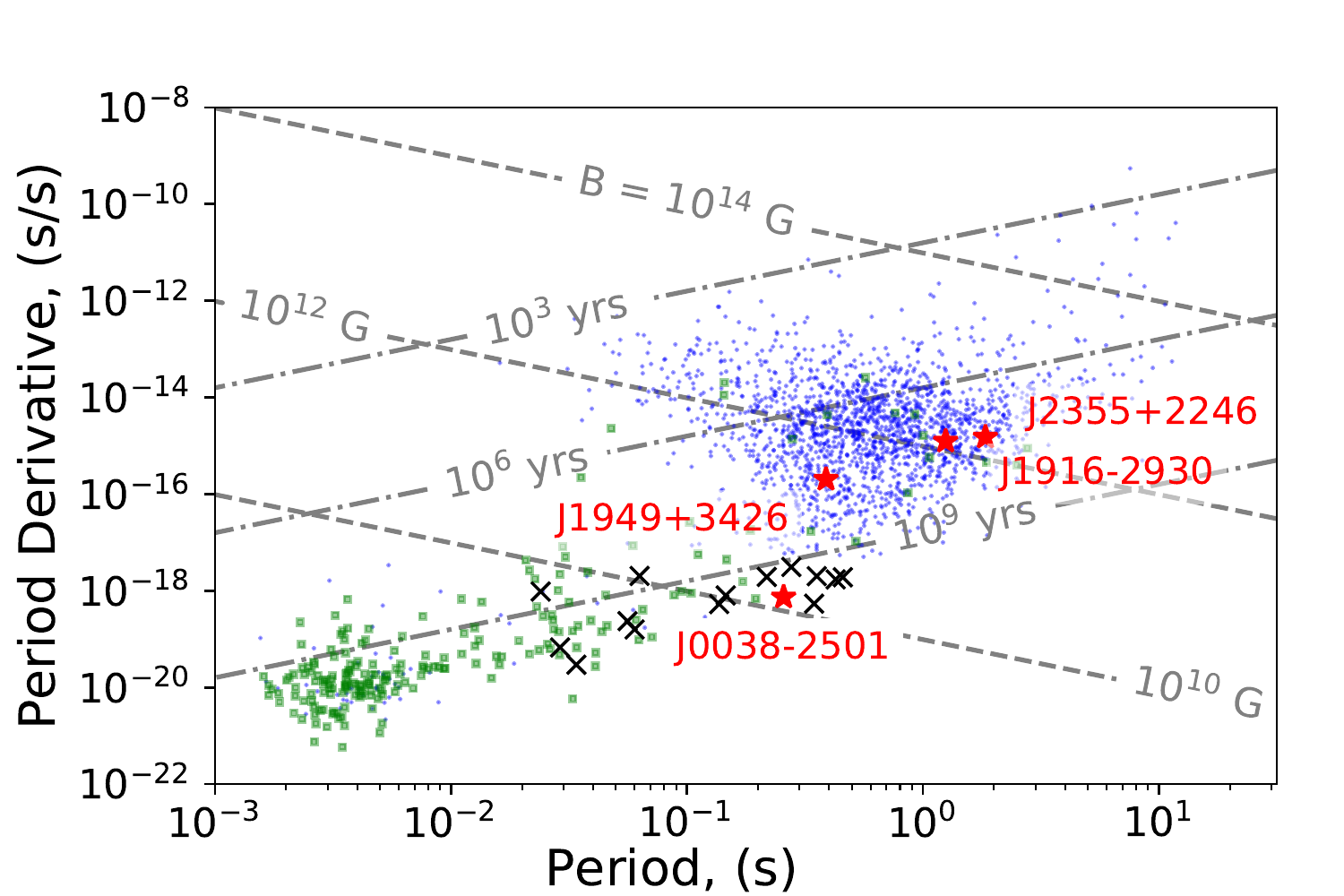}
	\caption{$P$ vs.\ $\dot{P}$ for the four pulsars discussed here (red stars, labeled) relative to known isolated pulsars (blue dots) and binary pulsars (green squares) from the ATNF Pulsar Catalog, Version 1.58 \citep{Manchester2005,Manchester2016}.  The disrupted recycled pulsar (DRPs) from \cite{Gotthelf2013} are shown as black $\times$ symbols.  Constant characteristic age in years is shown by grey dot-dashed lines.  Constant inferred surface dipole magnetic field is shown by grey dashed lines.}
\label{fig:ppdot}
\end{figure}

The locations on a $P-\dot P$ diagram are shown in Figure~\ref{fig:ppdot}.  Table~\ref{tab:spin} also includes the DM and calculated distance to each pulsar as well as the characteristic age, and calculated pseudo-luminosity for each pulsar.  The  flux density $S_{350}$ and pseudo luminosity $L_{350}=S_{350}\times d^2$ values reported in Table~\ref{tab:spin} were calculated for each of the timed pulsars using the signal-to-noise (S/N) from discovery observations as in \cite{stovall14}, and the search sensitivity \citep[][]{lorimer12}
\begin{equation}
S_{\rm min} = \frac{({\rm S/N})\,T_{\rm sys}}{G\sqrt{n_{\rm p}\,t_{\rm int}\,\Delta f}}\,\sqrt{\frac{W}{P-W}},
\label{eq:radiometer}
\end{equation}
where $T_{\rm sys}$ is the system temperature as listed in Table 1, $G = 2\,{\rm K\,Jy}^{-1}$ is the telescope gain, $n_{\rm p} = 2$ is the number of polarizations summed, $t_{\rm int} = 120\,$s is the integration time, $\Delta f = 80$\,MHz is the effective bandwidth, $W$ is the width of the pulse as detected by the system, and $P$ is the spin period.  $T_{\rm sys}$ includes sky temperatures $T_{\rm sky}$  listed in Table 1 as determined for the direction of each pulsar using the global sky model of  \citet{Oliveira2008} calculated at 350\,MHz.  

\section{Notes on Individual Pulsars}\label{sec:psrs}
\subsection{PSR~\psra}
As can be seen in Figure \ref{fig:ppdot}, the $\dot{P}$ and $P$ for PSR~\psra\ are low compared to typical non-recycled pulsars, implying a relatively low surface magnetic field strength ($B_{\rm s}$).  It was also the closest of the four timed pulsars at $600\,$pc (from YMW16; $320\,$pc from  NE2001).

Some pulsars with similar timing properties ($P>20$\,ms and $B_{\rm s}<3\times10^{10}$\,G) are described as disrupted recycled pulsars (DRPs; \citealt{Belczynski2013a}), where it is thought that the companion exploded in a supernova that unbound the system, stopping the recycling process and leaving the pulsar with intermediate properties between typical isolated pulsars and recycled MSPs \citep{Gotthelf2013}.  However, there is overlap between the properties of more traditional isolated pulsars and the DRPs.  \citet{Belczynski2013a} estimates that 0.3\% of isolated non-recycled pulsars may have $\dot P$ and $B_{\rm s}$ values similar to DRPs, which amounts to $\sim4$ pulsars out of the total population compared to 12 DRP considered in that paper.

\begin{figure}
\includegraphics[width=1.05\columnwidth]{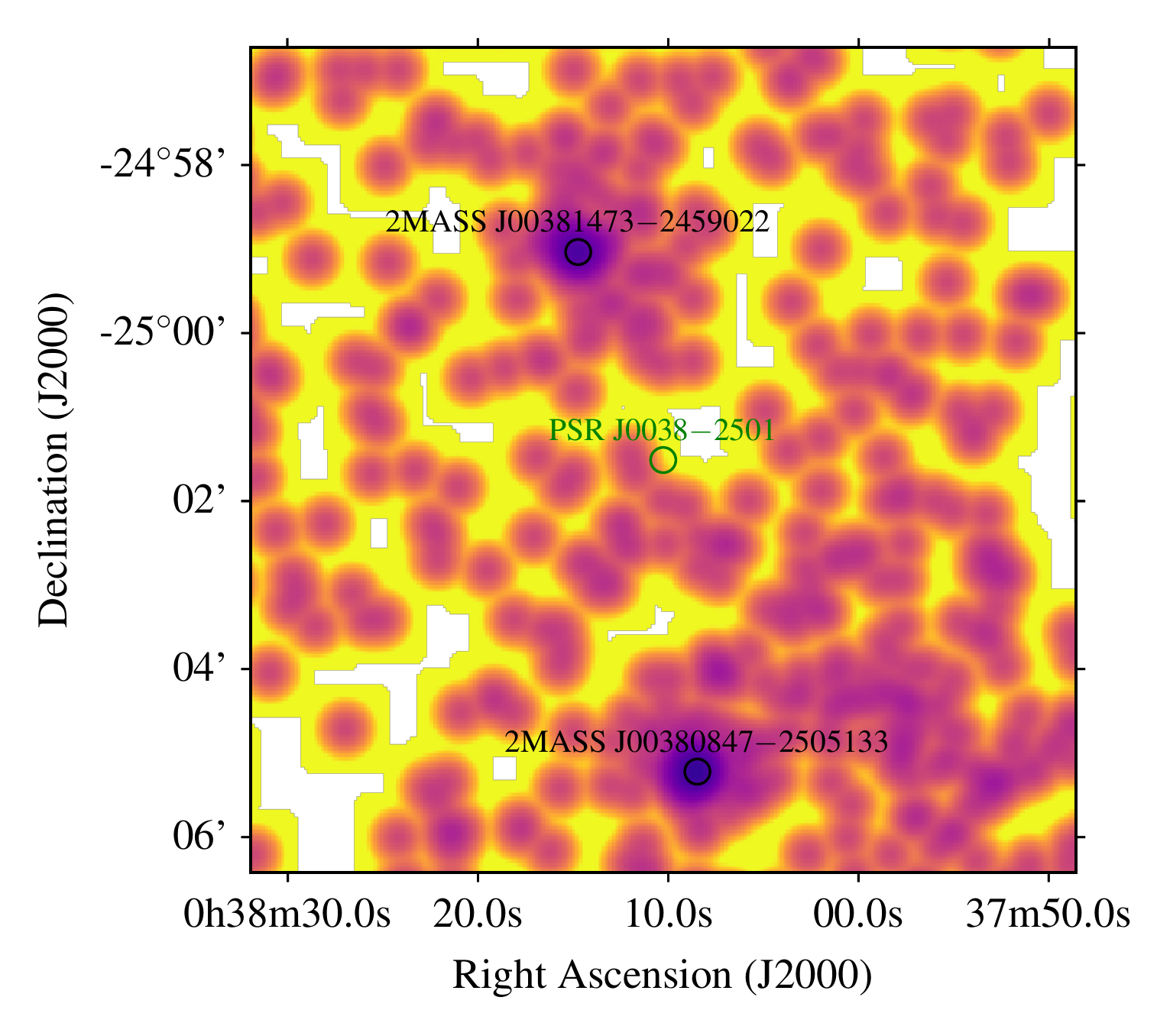}
	\caption{\swift\ XRT observation from 2009~November~16 centered on the position of PSR~\psra.  The image has been smoothed to have a full-width at half-maximum comparable to the XRT resolution of $18\arcsec$.  We show circles at the position of PSR~\psra\ (green) and two background point sources (black) from the Two Micron All Sky Survey \citep{2006AJ....131.1163S} which we used to verify the astrometry; the circle diameters are $18\arcsec$.}
\label{fig:swift}
\end{figure}

\begin{figure}
\includegraphics[width=1.1\columnwidth]{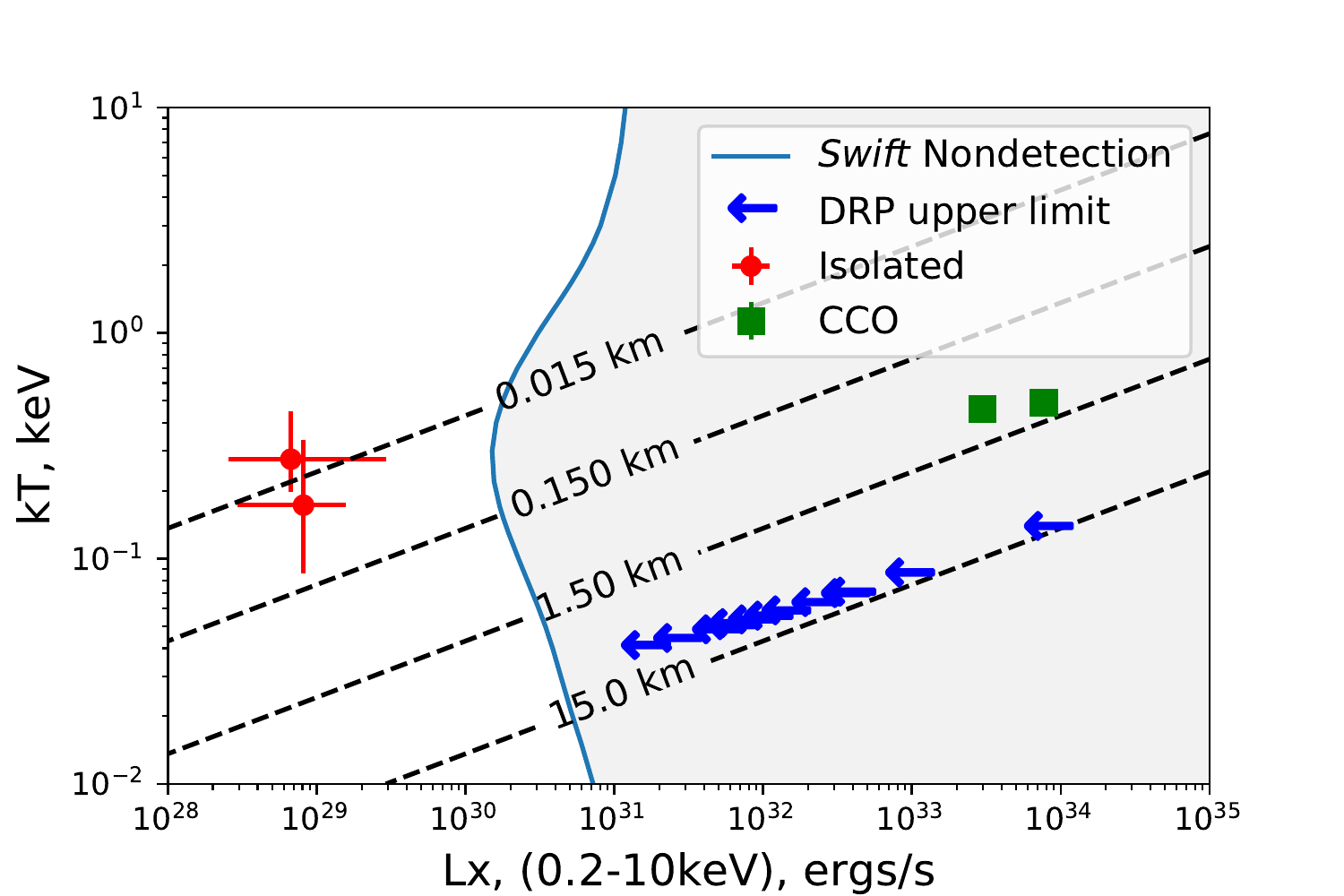}
	\caption{CCO and isolated pulsar X-ray luminosities and temperatures compared to limiting luminosities for non-detections for the 6\,ks\ \swift\ observation considered above.  The solid curve shows the luminosity (0.2--10\,keV) upper limit for non-detection of PSR~\psra, excluding the grey shaded area.  The limiting flux was calculated using \texttt{PIMMS} v4.8e\citep{Mukai1993} assuming $< 3$ counts in 6\,ks and column density $N_{\rm H}$ of $1.72\times10^{20}\,{\rm cm}^{-2}$ determined from the DM for \psra using the empirical relationship from \citet{He2013}.  The flux was calculated at various temperatures, $kT$, and converted to a luminosity assuming a distance of 600\,pc.  Dashed lines represent  equivalent source radii for blackbody models.  DRPs not detected in similar X-ray observations from \citet{Gotthelf2013} and \citet{Luo2015} are shown as blue arrows.  Luminosities for CCO observations from \citet{Halpern2010} are shown as green squares and for old isolated pulsars B0834+06 (3\,Myr) and B1133+16 (5\,Myr) from \citet{Gil2008} as red circles.}
\label{fig:Swift_L}
\end{figure}

An alternative explanation for the low magnetic field properties of PSR~\psra\ is that it could be an orphaned central compact object (CCO).  CCOs are young pulsars with low magnetic fields  that are found within or near supernova remnants (SNRs; \citealt{Gotthelf2013}).  The characteristic ages of CCOs calculated from their spin-down rates do not match the known ages of their associated SNRs.  It is unclear how CCOs evolve after they are formed.  \citet{Gotthelf2013} proposed that CCO descendants may have similar timing properties to the DRP pulsars, but would be expected to be younger and therefore may have visible thermal X-ray emission for up to $\sim10^5$\,yrs. \citet{Gotthelf2013} and \citet{Luo2015} searched known pulsars classified as DRPs (using same criteria as in \citealt{Belczynski2013a}) for X-ray emission with no detections.  We searched archival observations and found a 6\,ks \textit{Neil Gehrels Swift Observatory} (\swift) observation from 2009~November~16 that included the location of PSR~\psra.  Figure \ref{fig:swift} shows the \swift\ observation, which has 0\,counts in a circle of radius $18\arcsec$, consistent with the background expectation of $1.4\pm1.2\,$counts determined from the background rate of $0.0014\pm0.00005\,{\rm counts\,arcsec}^{-2}$.  Using \citet{1986ApJ...303..336G} we set an upper limit of $< 3$\,counts (95\% confidence). Figure~\ref{fig:Swift_L} compares the X-ray luminosities of young CCOs and old isolated pulsars with the upper limit for the \swift\ observation as well as upper limits for candidate DRPs from \citet{Gotthelf2013} and \citet{Luo2015}.  The failure to detect sources for these observations suggests that PSR~\psra\ and the candidate DRPs are not young ($\sim10^5$\,yrs old) orphaned CCOs.  PSR~\psra\ is the fifth potential DRP to be discovered in the GBNCC survey after PSRs J0358+6627, J0557$-$2948, J1434+7257, and J2122+5434 \citep{stovall14,lynch18,Kawash2018}.

The low $\dot{P}$ and low estimated distance for PSR~\psra\ also place limits on its proper motion and transverse velocity.  The observed period derivative for pulsars includes a component associated with proper motion determined by the Shklovskii effect \citep{Shklovskii1970}, calculated with
\begin{equation}
\dot{P_{\rm S}} = \frac{P\,V_{\rm T}^2}{c\,d},
\label{eq:shklov}
\end{equation}
where $V_{\rm T}$ is the transverse velocity, $c$ is the speed of light and $d$ is distance to the pulsar.  Additional period derivative components are present due to Galactic acceleration effects as described in \citet{Nice1995}, but these components were found to be less than 7\% of the measured period derivative and therefore were not included for this estimate.  Using this calculation and assuming that the intrinsic $\dot P$ is greater than 0 (i.e., the pulsar is spinning down), we limit $V_{\rm T}$ to  $<130\,{\rm km\,s}^{-1}$ for PSR~\psra\ (from YMW16; $<90\,{\rm km\,s}^{-1}$ from NE2001).  A relatively low transverse velocity is consistent with the lower expected natal kick velocities for DRPs \citep{Belczynski2013a}, (3-D velocity dispersion of 170\,${\rm km\,s}^{-1}$ compared to 265\,${\rm km\,s}^{-1}$ for isolated pulsars). 

\subsection{PSR~\psrc}
PSR~\psrc\ has a dispersion measure of 228.0\,\dmu\, which makes it the farthest of the four timed pulsars and among the top 7\% most distant Galactic field pulsars. The distance determined from the YMW16 model is 12.3\,kpc (9.8\,kpc from NE2001 model).  The pulse profile shown in Figure \ref{fig:profiles} has a tail resembling profiles that exhibit scatter broadening as described in \citet{Bhat2004} which would be consistent with a large distance.  We fit a one-sided exponential function to the pulse profile after the peak in four sub-bands centered at 313, 338, 358, and 389\,MHz. We find decay times of 30.1$\pm{0.8}$, 20.1$\pm{0.4}$, 15.7$\pm{0.5}$, and 15.1$\pm{0.7}$\,ms, although we note that this does not account for any intrinsic pulse width or frequency evolution of the pulse shape.  The timescale decreases with increasing frequency as expected for interstellar pulse broadening, but it should be noted that the range of frequencies was relatively narrow for this evaluation.  The NE2001 model predicts a much lower pulse broadening timescale of $< 1\,$ms at 350\,MHz, while the YMW16 model predicts a larger timescale of $60\,$ms.  Additional observations at a higher frequency could help confirm whether the profile shape is impacted by pulse broadening.  The large distance for PSR~\psrc\ suggests that it must be  relatively bright to have been detected in the GBNCC search. The pseudo-luminosity (calculated as described in Section 3) is one of the highest reported for pulsars discovered by GBNCC \citep{stovall14}.  
\subsection{PSR~\psrb}
PSR J1916$-$2939 has properties typical for isolated non-recycled pulsars with a longer, 1.84\,s period and large period derivative due to a relatively high surface magnetic field.  

\subsection{PSR~\psrd}
PSR~\psrd also has properties typical for young, isolated non-recycled pulsars.  The signal-to-noise ratio in the observations was relatively low at $\sim12$.  Several TOAs were removed after confirming that no significant pulse profile was visible.  Some evidence of pulse nulling \citep[where the pulsar appears to turn off for a some numbers of pulses;][]{1970Natur.228...42B} was noted for roughly 30\% of the pulses in a 2 minute discovery observation. We reviewed additional timing observations for nulling behavior, but were unable to confirm this behavior because of excess RFI.  Additional observations can confirm whether this pulsar nulls or not.

\section{Conclusions}
In this paper, we report the timing solutions for four pulsars discovered in the GBNCC survey.  The properties of the timed pulsars are varied indicating differing evolutionary paths, which supports one of the GBNCC objectives of characterizing the pulsar population to better understand the underlying physical phenomena.  PSR~\psra was found to have an unusually low magnetic field suggesting that it may be a DRP or possibly an orphaned CCO.  An archival \swift\ X-ray observation did not find a source at the location, suggesting that PSR~\psra\ is not a young orphaned CCO, but it could be an older source.  Additional observations are suggested to determine the proper motion of PSR~\psra\ which may help distinguish between evolutionary models.  The farthest of the four pulsars according to the DM-distance models reported here was PSR~\psrc.  Calculations indicate that it \textbf{may be} one of the highest pseudo-luminosity pulsars discovered in the GBNCC survey.  The profile may show evidence of pulse broadening.  Observations at higher frequency would allow better evaluation of the intrinsic pulse profile and  determination of the extent of scattering.

\acknowledgements
The undergraduate students would like to thank the University of Wisconsin--Milwaukee for providing this research opportunity through the First Year Research Experience (FYRE) initiative.
We thank the referee for their time and comments.
The Green Bank Observatory is a facility of the National Science Foundation operated under cooperative agreement by Associated Universities, Inc.  Support was provided by the NANOGrav NSF Physics Frontiers Center award number 1430284.
Pulsar research at UBC is supported by an NSERC Discovery Grant and by the Canadian Institute for Advanced Research.
This research has made use of data obtained through the High Energy Astrophysics Science Archive Research Center Online Service, provided by the NASA/Goddard Space Flight Center.  JvL acknowledges funding from the European Research Council grant n. 617199.

\facility{GBT, Swift}

\software{GSM \citep{Oliveira2010}, PRESTO (\url{http://www.cv.nrao.edu/~sransom/presto/}),
PSRCHIVE \citep{hotan04,vanStraten2012}, TEMPO2 \citep{hobbs06}),
}

\end{document}